\documentclass[journal]{IEEEtran}
\IEEEoverridecommandlockouts
\usepackage{cite}
\usepackage{amsmath,amssymb,amsfonts,amsthm}
\usepackage{comment}
\newcommand{\norm}[1]{\left\lVert#1\right\rVert}
\usepackage{algorithmic}
\usepackage{algorithm}
\usepackage{enumerate}  
\DeclareMathOperator{\Tr}{Tr}
\DeclareMathOperator{\diag}{diag}
\DeclareMathOperator{\vect}{vec}
\usepackage[english]{babel}

\newtheorem{remark}{Remark}

\usepackage{subcaption}
\usepackage{graphicx}
\usepackage{textcomp}
\usepackage{optidef}
\usepackage{xcolor}

\begin{document}

\title{Integrated Sensing, Computing and Semantic Communication for Vehicular Networks % 
\thanks{Copyright (c) 20xx IEEE. Personal use of this material is permitted. However, permission to use this material for any other purposes must be obtained from the IEEE by sending a request to pubs-permissions@ieee.org.}}

\author{Yinchao Yang, Zhaohui Yang, Chongwen Huang, Wei Xu, Zhaoyang Zhang,\\
Dusit Niyato, \IEEEmembership{Fellow, IEEE}, and Mohammad Shikh-Bahaei, \IEEEmembership{Senior Member, IEEE}
%\thanks{This research is supported in part by National Key R\&D Program of China (Grant No. 2023YFB2904804), Young Elite Scientists Sponsorship Program by CAST 2023QNRC001,  Zhejiang Key R\&D Program under Grant 2023C01021, the Fundamental Research Funds for the Central Universities, K2023QA0AL02. supported by the National Research Foundation, Singapore and Infocomm Media Development Authority under its Future Communications Research \& Development Programme (FCP-NTU-RG-2022-010 and FCP-ASTAR-TG-2022-003), Singapore Ministry of Education (MOE) Tier 1 (RG87/22 and RG24/24), the NTU Centre for Computational Technologies in Finance (NTU-CCTF), and the RIE2025 Industry Alignment Fund - Industry Collaboration Projects (IAF-ICP) (Award I2301E0026), administered by A*STAR, as well as supported by Alibaba Group and NTU Singapore through Alibaba-NTU Global e-Sustainability CorpLab (ANGEL). This research is also supported in part by National Key R\&D Program of China (Grant No. 2023YFB2904804), Young Elite Scientists Sponsorship Program by CAST 2023QNRC001,  Zhejiang Key R\&D Program under Grant 2023C01021, the Fundamental Research Funds for the Central Universities, K2023QA0AL02.}
\thanks{Yinchao Yang and Mohammad Shikh-Bahaei are with the Department of Engineering, King's College London, London, UK (emails: yinchao.yang@kcl.ac.uk; m.sbahaei@kcl.ac.uk).}
\thanks{Wei Xu is with the National Mobile Communications Research Laboratory, Southeast University, Nanjing 210096, China, and also with the Purple Mountain Laboratories, Nanjing 211111, China (email: wxu@seu.edu.cn).}
\thanks{Zhaohui Yang, Chongwen Huang and Zhangyang Zhang are with the College of Information Science and Electronic Engineering, Zhejiang University, Hangzhou, Zhejiang 310027, China, and Zhejiang Provincial Key Lab of Information Processing, Communication and Networking (IPCAN), Hangzhou, Zhejiang, 310007, China (emails: yang\_zhaohui@zju.edu.cn; chongwenhuang@zju.edu.cn; ning\_ming@zju.edu.cn).}
\thanks{Dusit Niyato is with the College of Computing and Data Science, Nanyang Technological University, Singapore 639798, Singapore (email: dniyato@ntu.edu.sg).}
}

\maketitle

\begin{abstract} 
This paper introduces a novel framework for integrated sensing, computing, and semantic communication (ISCSC) within vehicular networks comprising a roadside unit (RSU) and multiple autonomous vehicles. Both the RSU and the vehicles are equipped with local knowledge bases to facilitate semantic communication. The framework incorporates a secure communication design to ensure that messages intended for specific vehicles are protected against interception. Within this model, an extended Kalman filter (EKF) is employed by the RSU to track all vehicles accurately. We formulate a joint optimisation problem that balances maximising the probabilistic-constrained semantic secrecy rate for each vehicle while minimising the sum of the posterior Cramer-Rao bound (PCRB), constrained by the RSU’s computing capabilities. This non-convex optimisation problem is addressed using Bernstein-type inequality (BTI) and alternating optimisation (AO) techniques. Simulation results validate the effectiveness of the proposed framework, highlighting its advantages in reliable sensing, high data throughput, and secure communication.
\end{abstract}

\begin{IEEEkeywords}
Integrated sensing and communication, transmit beamforming, semantic communication, vehicular networks, and physical layer security.
\end{IEEEkeywords}

\IEEEpeerreviewmaketitle

\section{Introduction}

\IEEEPARstart{R}{ecently}, integrated sensing and communication (ISAC) has been considered one of the vital techniques for vehicular networks \cite{su2022secure}. The dual functionalities effectively reduce the usage of radio frequency and potentially improve security, because the sensing signals can act as artificial noise on top of the communication signals. Yuan \textit{et al.} \cite{yuan2020bayesian} explored ISAC beamforming design in vehicular networks, aiming to accurately track vehicles while maintaining conventional communication with them. Besides, Dong \textit{et al.} \cite{dong2022sensing} focused on ISAC resource allocation design in vehicular networks, with a similar objective of precise vehicle tracking and communication. Considering the imperfect channel, Jia \textit{et al.} \cite{jia2023physical} investigated robust ISAC beamforming design in vehicular networks, aiming to mitigate the effects of channel errors when communicating with communication users (CUs) and detecting the targets.

With the rapid development of artificial intelligence, semantic communication has also been proven to be an effective technology for addressing resource scarcity. A breakthrough in semantic communication is that it transcends Shannon's paradigm, as opposed to conventional communication, which is constrained by Shannon's capacity limit \cite{sun2024s}. Rather than transmitting the entire message, semantic communication focuses on extracting and transmitting just its meaning. Several semantic coding strategies have been proposed for different data formats. Su \textit{et al.} \cite{su2023semantic} proposed a resource allocation scheme for semantic communication in vehicular networks, where the proposed scheme is robust against high-dynamic vehicles. Besides, Xia \textit{et al.} \cite{xia2023xurllc} addressed the knowledge base construction and vehicle service pairing for a semantic communication-enabled vehicular network.

While semantic communication optimizes data transmission, its potential can be further unlocked by integrating it with sensing technologies. The integration of sensing and semantic communication enhances vehicular networks by improving efficiency, reliability, and real-time decision-making. Sensing provides context-awareness, optimizing semantic information extraction and transmission for low-latency and high-reliability communication. By prioritizing critical events from environmental data, semantic communication enhances collision avoidance, traffic optimization, and situational awareness. This synergy reduces spectrum congestion, streamlines message transmission, and maximizes resource efficiency, making vehicular networks more adaptive, intelligent, and resilient.

Different from the existing works \cite{yuan2020bayesian, dong2022sensing, jia2023physical, su2023semantic, xia2023xurllc}, in this paper, we explore the integration of ISAC and semantic communication in a vehicular network while considering the computing ability of the roadside unit (RSU). We name this framework integrated sensing, computing and semantic communication (ISCSC). The proposed framework allows a higher transmission rate without violating the sensing performance originally provided by ISAC. Additionally, semantic communication requires each party to have their local knowledge base (KB) to decode the semantic messages. In this regard, if the unintended receivers do not have the same KB as the RSU, they are not able to fully decode the intercepted semantic messages. In such a way, the communication security is enhanced. The contribution of this paper is summarised as follows:
\begin{enumerate}
    \item For the first time, this work presents an analysis of the ISCSC framework within vehicular networks. In this framework, the RSU is responsible for processing and transmitting semantic messages to designated vehicles while ensuring secure communication by preventing unauthorised access from unintended vehicles. At the same time, the RSU continuously monitors and tracks the real-time locations of all vehicles within the network.
    \item We formulate an outage probability-constrained optimisation problem to handle the errors brought by the channel prediction. To solve the formulated non-convex optimisation problem, an efficient algorithm is proposed based on Bernstein-Type Inequality and alternating optimisation. 
\end{enumerate}

%\vspace{-0.4cm}

\section{System Model}
\begin{figure}[!t]
    \centering
    \includegraphics[width = 0.65\linewidth]{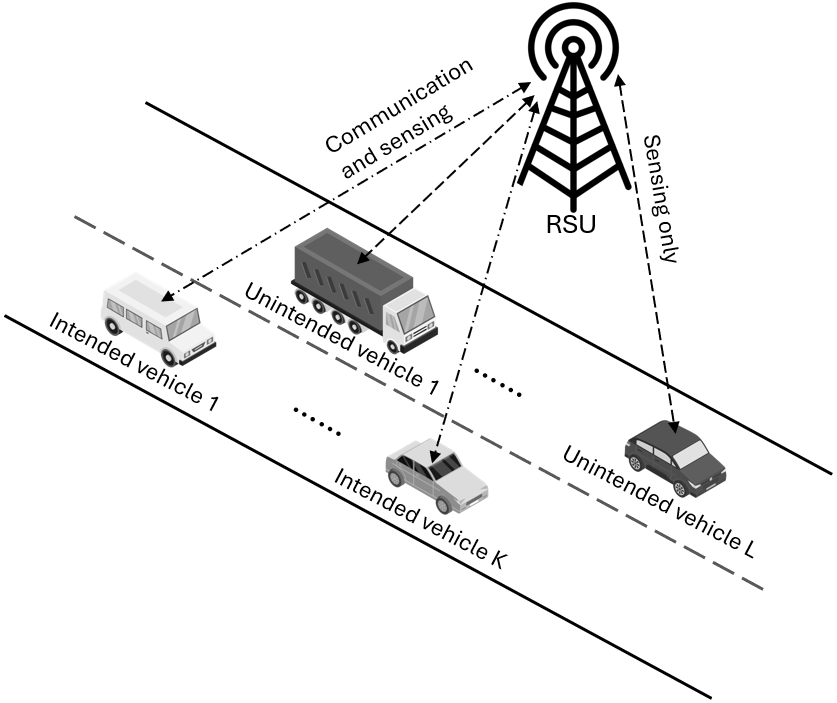}
    \caption{System model of ISCSC in a vehicular network.}
    \label{response 1}
\end{figure}

We consider the design of an ISCSC system where an RSU is equipped with a uniform linear array (ULA) of $N$ antennas. Let $L$ and $K$ denote the sets of unintended vehicles and intended vehicles, respectively, for communication. Each vehicle $i \in L \cup K$ is equipped with a single antenna. The RSU aims to track all vehicles in the system. Meanwhile, the RSU communicates with all the intended vehicles using semantic messages. Similar to \cite{liu2020radar, dong2022sensing}, we consider that vehicles maintain a constant velocity and travel in the same direction, e.g., vehicles on a highway. In contrast to  \cite{liu2020radar, dong2022sensing}, our scenario involves vehicles travelling along a double-lane straight road parallel to the RSU. An illustration of the system setting is shown in Fig. \ref{response 1}. Since the vehicles are in motion, the subscript $t$ will be used to denote the time slot $t$ throughout the remainder of this paper.

%\vspace{-0.4cm}

\subsection{Communication and Sensing Models}
The received signal at the vehicle $k$ can be characterised by
\begin{equation}\label{eq1}
y_{k,t} = \mathbf{h}_{k,t}^H \mathbf{x}_t + n_k,
\end{equation}
where $\mathbf{h}_{k,t}\in\mathbb{C}^{N\times1}$ is the channel vector, and the superscript $H$ means the Hermitian transpose of a matrix. Additionally, $\mathbf{x}_t\in\mathbb{C}^{N\times1}$ is the transmitted signal, and $n_k \sim \mathcal{CN}(0,\sigma^2_c)$ is the additive Gaussian noise. The echo signal received by the RSU about vehicle $i$ after applying a matched filter can be formulated by
\begin{equation}\label{eq4}
    \mathbf{\hat{y}}_{i,t} = \beta_{i,t} \mathbf{a}(\theta_{i,t}) \mathbf{a}^H(\theta_{i,t}) \mathbf{x}_t + \mathbf{n}_i,
\end{equation}
where $\mathbf{a}(\theta_{i,t}) \in \mathbb{C}^{N\times1}$ is the steering vector, $\beta_{i,t}$ is the round-trip path loss coefficient, and $\mathbf{n}_i \in \mathbb{C}^{N\times1}$ is additive Gaussian noise vector with zero mean and variance of $\sigma^2_r \mathbf{I}_{N \times 1}$, where $\mathbf{I}$ is an identity matrix. 

The joint sensing and semantic communication signal $\mathbf{x}_t$ transmitted by the RSU is given by
\begin{equation}\label{eq5}
    \mathbf{x}_t = \sum_{k \in K} \mathbf{w}_{k,t} c_{k,t} +\sum_{i \in L \cup K} \mathbf{r}_{i,t} z_{i,t},
\end{equation}
where $\mathbf{w}_{k,t} \in \mathbb{C}^{N \times 1}$ and $\mathbf{r}_{i,t} \in \mathbb{C}^{N \times 1}$ represent the beamforming vectors for communication and sensing, respectively. Additionally, $c_{k,t}$ and $z_{i,t}$ denote the semantic message and sensing signal, respectively.

%$\mathbf{W}_t \in \mathbb{C}^{N \times K}$ denotes the precoding matrix and $\mathbf{c}_t \in \mathbb{C}^{K\times1}$ denotes the semantic message. Moreover, $\mathbf{R}_t \in \mathbb{C}^{N \times (L + K)}$ represents the radar beamforming matrix and $\mathbf{z} \in \mathbb{C}_t^{(L + K)\times1}$ is the sensing signal. 
The covariance matrix of the transmit waveform can be derived as follows:
\begin{equation}\label{eq9}
    \mathbf{R}_{x_t} = \mathbb{E}[\mathbf{x}_t \mathbf{x}_t^H] = \sum_{k \in K } \mathbf{W}_{k,t} + \sum_{i \in L \cup K} \mathbf{R}_{i,t},
\end{equation}
where $\mathbf{W}_{k,t} = \mathbf{w}_{k,t} \mathbf{w}_{k,t}^H$, $\mathbf{R}_{i,t}= \mathbf{r}_{i,t} \mathbf{r}_{i,t}^H$, and $\mathbb{E}[\cdot]$ represents the expected value. 

%\vspace{-0.44cm}

\subsection{State Evolution Model and Extended Kalman Filter}
To model the mobility of vehicles, we adopt the state model. The state model describes the correlation between two successive samples in the time domain. The state model for each vehicle $i$ is presented as follows:
\begin{equation}\label{state model}
    \begin{aligned}
        & \theta_{i, t} = \theta_{i, t-1} + d^{-1}_{i, t-1} v_{i, t-1}\Delta T \sin{\theta_{i, t-1} + u_\theta},\\
        & d_{i, t} = d_{i, t-1} - v_{i, t-1}\Delta T \cos{\theta_{i, t-1}} + u_d,\\
        & v_{i, t} = v_{i, t-1} + u_v,\\
        & \beta_{i, t} = \beta_{i, t-1} \left(1 + d^{-1}_{i, t-1} v_{i, t-1} \Delta T \cos \left(\theta_{i, t-1} \right) \right) + u_\beta,
    \end{aligned}
\end{equation}
where $\mathbf{q}_{i,t} = [\theta_{i,t}, d_{i,t}, v_{i,t}, \beta_{i,t}]$ denotes the vector with state variables (i.e., angle, distance, velocity and path loss) for vehicle $i$ at time slot $t$. $\Delta T$ is the length of a time slot, and $t-1$ means the previous time slot. Lastly, $\mathbf{u}_t = [u_\theta, u_d, u_v, u_\beta]$ are the state noise. The observed echo signal depends on the current state variables $\mathbf{q}_{i,t}$. Therefore, we can conclude the state model and the observation model as:
\begin{equation}\label{eq55}
    \begin{cases}
      \text{State Model:} & \mathbf{q}_{i,t} = \mathbf{g}_1(\mathbf{q}_{i,t-1}) + \mathbf{u}_t,\\
      \text{Observation Model:} & \mathbf{r}_{i,t} = \mathbf{g}_2(\mathbf{q}_{i,t}) + \mathbf{e}_t,\\
    \end{cases}
\end{equation}
where $\mathbf{g}_1(\cdot)$ and $\mathbf{g}_2(\cdot)$ are non-linear functions. $\mathbf{u}_t$ and $\mathbf{e}_t$ are noise vectors with zero-mean Gaussian distribution, and their covariance matrices can be formulated by
\begin{equation}\label{eq56}
       \mathbf{Q}_1 = \diag \left(\sigma^2_\theta, \sigma^2_d, \sigma^2_v, \sigma_\beta^2 \right), \mathbf{Q}_2 = \diag \left(\sigma^2_e \mathbf{1}_{1 \times 2}, \sigma^2_{\hat{d}}, \sigma^2_{\hat{v}} \right).
\end{equation}
where $\mathbf{1}$ is an all one matrix. To provide accurate beam tracking of the vehicles during long-term moving, an extended Kalman filter (EKF) is applied for angle prediction. The EKF is effective for nonlinear systems by linearizing the state model, making it computationally efficient and scalable for real-time, high-dimensional applications. In contrast, other filters, such as particle filter (PF), though more flexible for highly nonlinear cases, require a large number of particles to maintain accuracy, leading to higher computational costs and potential sample degeneracy issues. The detailed procedures of EKF are provided in \cite{kay1993fundamentals}, we briefly describe them below.

\begin{enumerate}[(i)]
    \item Available observations at $t-1$: At time $ t $, we assume that the state variables $\mathbf{q}_{i, t-1}$ from the previous time step are known.

    \item State prediction at \( t \): Before transmitting any sensing signals, we estimate the state variables using past information. This is achieved via \eqref{state model}. Using the estimated state, we compute the predicted channel via
    $ \Bar{\mathbf{h}}_{i,t} = \hat{\beta}_{i, t|t-1} \mathbf{a}\left(\hat{\theta}_{i, t|t-1}\right)$.

    \item Linearization: To account for the non-linearities in the state transition and measurement functions, we perform a first-order Taylor expansion and compute the Jacobian matrices. We use $ \mathbf{G}_1 $ to represent the Jacobian of $ \mathbf{g}_1(\cdot) $ and $ \mathbf{G}_2 $ to represent the Jacobian of \( \mathbf{g}_2(\cdot) \). These Jacobian matrices capture the sensitivity of the state and measurement functions with respect to the state variables.

    \item Mean square error (MSE) prediction:
    To assess the uncertainty in the state prediction, we compute the MSE matrix via $
        \mathbf{M}_{i, t|t-1}= \mathbf{G}_{1} \mathbf{M}_{i, t-1} \mathbf{G}_{1}^H + \mathbf{Q}_1$.
    Here, \( \mathbf{M}_{i, t-1} \) is the MSE matrix from the previous time step (assume this is available) and \( \mathbf{Q}_1 \) is the process noise covariance matrix, modelling system uncertainties.

    \item Kalman gain calculation:
    Once the signal is transmitted and the echo is received, we update the state using the measurement. The Kalman Gain optimally balances the contribution of the new measurement relative to the prediction, and it can be calculated by
        $\mathbf{K}_{i,t} = \mathbf{M}_{i, t|t-1} \mathbf{G}_{2}^H (\mathbf{Q}_2 + \mathbf{G}_{2} \mathbf{M}_{i, t|t-1} \mathbf{G}_{2}^H)^{-1}$. Here, \( \mathbf{Q}_2 \) is the measurement noise covariance matrix and \( \mathbf{G}_2 \) is the Jacobian matrix capturing measurement sensitivity. To obtain \( \mathbf{G}_2 \), we differentiate (2) with respect to each observed state variable.

    \item State tracking:
    Using the actual measurement, the predicted state, and the computed Kalman Gain, we refine the state estimate via
    $\mathbf{\hat{q}}_{i,t} = \mathbf{\hat{q}}_{i, t|t-1} + \mathbf{K}_{i,t} \big(\mathbf{r}_{i,t} - \mathbf{g}_2(\mathbf{\hat{q}}_{i, t|t-1}) \big)$. Here, \( \mathbf{r}_{i,t} \) is the actual measurement and \( \mathbf{g}_2(\mathbf{\hat{q}}_{i, t|t-1}) \) represents the predicted measurement.

    \item MSE matrix update:
    Following the state update, we refine the MSE matrix via
    $\mathbf{M}_{i,t} = \left (\mathbf{I} - \mathbf{K}_{i,t} \mathbf{G}_{2} \right) \mathbf{M}_{i, t|t-1}$.

    \item Transition to the next time step \( t+1 \):
    The system advances to the next time step \( t+1 \). The updated state estimate at time \( t \) serves as the prior knowledge for the next prediction cycle, and its formulation is given by $\mathbf{\hat{q}}_{i, t+1|t} = \mathbf{g}_1(\mathbf{\hat{q}}_{i, t})$.
\end{enumerate}

This completes the state tracking process for time \( t \), and the procedure is repeated recursively in subsequent time slots.

%\vspace{-0.4cm}
\section{Performance Indicators}

%\vspace{-0.1cm}
\subsection{Semantic Communication}
The semantic transmission rate is defined as the number of bits received by the vehicle following the extraction of semantic information, and the formulation can be given by \cite{yang2024secure}
\begin{equation}\label{eq11}
    S_{k,t} = \frac{\iota}{\rho_{k,t}} \log \left(1+\gamma_{k,t}\right),
\end{equation}
where $0\leq \rho_{k,t} \leq 1$ represents the semantic extraction ratio for vehicle $k$ at time slot $t$, and $\iota$ is a scalar value that represents the word-to-bit ratio. The term $\gamma_{k,t}$ denotes the signal-to-interference-plus-noise ratio (SINR) for the $k$-th vehicle, expressed as:
\begin{equation}\label{eq12}
\footnotesize
    \gamma_{k,t}= \frac{\Tr \left(\mathbf{h}_{k,t} \mathbf{h}_{k,t}^H \mathbf{W}_{k,t} \right)}{\Tr \left(\mathbf{h}_{k,t} \mathbf{h}_{k,t}^H \left(\sum_{k'\in K, k' \neq k}\mathbf{W}_{k',t} + \sum_{i \in L \cup K} \mathbf{R}_{i,t} \right)\right) + \sigma^2_c},
\end{equation}
where $\Tr(\mathbf{A})$ represents the trace of matrix $\mathbf{A}$.

We have derived the lower bound for $ \rho_{k,t}$ as presented in \cite{yang2024secure}, which is given by:
\begin{equation}\label{eq14}
    \rho_{k,t} \geq \frac{1}{1 - \ln Q_t + \sum_{g=1}^G w_{g,k,t} \log p_{g,k,t}},
\end{equation}
where $Q_t$ is the global lower bound of all bilingual evaluation understudy scores (BLEU). Additionally, $w_{g,k,t}$ is the weight of the g-grams, $G$ is the total number of g-grams required to represent a sentence, and the precision score $p_{g,k,t}$ depends on each vehicle. To prevent data breaches, we evaluate the worst-case semantic secrecy rate (SSR) to assess the security level. A higher SSR means a lower chance of data breaches.

Before formulating this, it is necessary to define the SINR of the $l$-th unintended vehicle in relation to the $k$-th intended vehicle:
\begin{equation}\label{eq15}
\footnotesize
    \Gamma_{l | k, t}  = \frac{\Tr\left(\mathbf{h}_{l,t} \mathbf{h}_{l,t}^H \mathbf{W}_{k,t}\right) }{\Tr\left(\mathbf{h}_{l,t} \mathbf{h}_{l,t}^H \left(\sum_{k' \in K , k' \neq k} \mathbf{W}_{k',t} +   \sum_{i \in L \cup K} \mathbf{R}_{i,t}\right) \right)  +  \sigma^2_c}.
\end{equation}

In the worst-case scenario, where the unintended vehicle has an extensive KB similar to that of the RSU and the intended vehicle, the semantic transmission rate for the vehicle $l$ related to the vehicle $k$ is determined as follows:
\begin{equation}\label{eq16}
    S_{l | k, t} = \frac{\iota}{\rho_{k,t}} \log \left(1+\Gamma_{l | k,t} \right).
\end{equation}

In this way, the worst-case SSR of the $k$-th vehicle is formulated by
\begin{equation}\label{eq17}
    SSR_{k,t} = \min_{l \in L} [S_{k,t} - S_{l | k,t}]^+,
\end{equation}
where $[\cdot]^+$ means $\max(0, \cdot)$.

\begin{remark}
The semantic secrecy rate can be rewritten as \( SSR_k \propto \log \left( \frac{1+\gamma}{1+\Gamma} \right) \), where the numerator represents the quantity of the received message per second for the legitimate vehicle, and the denominator represents the unintended vehicle's. A high ratio indicates minimal information intercepted by the unintended vehicle, reducing the risk of data breaches. By taking the log scale, we conclude that higher SSR corresponds to a lower likelihood of data breaches, as the unintended vehicle’s received data rate approaches zero, making interception ineffective.
\end{remark}

\vspace{-0.5cm}

\subsection{Power Budget}
Deriving semantic information from a traditional message heavily relies on machine learning techniques. Hence, considering computational power as a component of the overall transmission power budget is crucial. In \cite{yang2024secure}, we employ a natural logarithm function for computing the computational power:
\begin{equation}\label{eq18}
    P^{\text{comp}}_{t} =   -F\sum_{k \in K}\ln \left(\rho_{k,t} \right),
\end{equation}
where $F$ is a coefficient that converts a magnitude to its power. On the other hand, the total communication and sensing energy consumption at the RSU side is given by
\begin{equation}\label{eq19}
     P^{\text{c\&s}}_{t} = \Tr \left (\sum_{k \in K} \mathbf{W}_{k,t} + \sum_{i \in L \cup K} \mathbf{R}_{i,t} \right).
\end{equation}

\vspace{-0.4cm}
\subsection{Posterior Cramér-Rao Bound}
For assessing static target sensing performance, the Cramér-Rao bound (CRB) offers a lower bound of the MSE and can be expressed in a closed form. Defining the parameters to be estimated as $\xi_{i,t} = [\theta_{i,t}, \beta_{i,t}]$, the Fisher information matrix (FIM) about $\xi_{i,t}$ is given by:
\begin{equation}\label{eq21}
    \mathbf{J}^o_{i,t} = 
    \begin{bmatrix}
        {J}_{\theta_{i,t} \theta_{i,t}} & \mathbf{J}_{\theta_{i,t} \beta_{i,t}}\\
        \mathbf{J}_{\theta_{i,t} \beta_{i,t}}^T & \mathbf{J}_{\beta_{i,t} \beta_{i,t}}
    \end{bmatrix},
\end{equation}
where 
\begin{equation}\label{eq22}
\footnotesize
    {J}_{\theta_{i,t} \theta_{i,t}} = \frac{2T|\beta_{i,t}|^2}{\sigma^2_r} \Tr \left(\dot{\mathbf{B}}_{\theta_{i,t}} \mathbf{R}_{x_t} \dot{\mathbf{B}}_{\theta_{i,t}}^H \right),
\end{equation}
\begin{equation}
\footnotesize
    \mathbf{J}_{\beta_{i,t} \beta_{i,t}} = \frac{2T}{\sigma^2_r} \Tr \left(\mathbf{B}_{i,t} \mathbf{R}_{x_t} \mathbf{B}_{i,t}^H \right) \mathbf{I},
\end{equation}
and
\begin{equation}\label{eq23}
\footnotesize
    \mathbf{J}_{\theta_{i,t} \beta_{i,t}} = \frac{2T \beta_{i,t}^*}{\sigma^2_r} \Re\{ \Tr\left(\mathbf{B}_{i,t} \mathbf{R}_{x_t} \dot{\mathbf{B}}^H_{\theta_{i,t}} \right)[1\;j]\}, 
\end{equation}
with $T$ representing the number of samples, $\mathbf{B}_{i,t} = \mathbf{a}(\theta_{i,t}) \mathbf{a}^H(\theta_{i,t})$, and $\Dot{\mathbf{B}}_{\theta_{i,t}} = \frac{\partial \mathbf{B}_{i,t}}{\partial \theta_{i,t}}$. In the context of moving vehicles, the CRB for parameter estimation must account for both the observation and state models, rendering the Posterior CRB (PCRB) a more robust sensing metric. The posterior FIM can be formulated by
\begin{equation}\label{eq63}
    \mathbf{J}^p_{i,t} = \mathbf{J}^o_{i,t} + \mathbf{J}^e_{i,t},
\end{equation}
with $\mathbf{J}^o_{i,t}$ being the FIM from the observation and $\mathbf{J}^e_{i,t}$ being the FIM extracted from the prior distribution information. According to \cite{liu2020radar}, $\mathbf{J}^e_{i,t}$ can be obtained as follows:
\begin{equation}\label{eq64}
        \mathbf{J}^e_{i,t} = \mathbf{M}_{i, t|t-1}^{-1} = \left(\mathbf{G}_{1} \mathbf{M}_{i, t-1} \mathbf{G}_{1}^H + \mathbf{Q}_1 \right)^{-1}.
\end{equation}

We only focus on the error of the estimated angles, therefore, $\mathbf{J}^p_{i,t}$ reduces to:
\begin{equation}\label{eq65}
    \mathbf{J}^p_{i,t} = \begin{bmatrix}
         {J}_{\theta_{i,t} \theta_{i,t}} + {\mathbf{M}_{i, t | t-1}^{-1}}_{[1,1]} & \mathbf{J}_{\theta_{i,t} \beta_{i,t}} \\
        \mathbf{J}_{\beta_{i,t} \theta_{i,t}}^T & \mathbf{J}_{\beta_{i,t} \beta_{i,t}}
    \end{bmatrix},
\end{equation}
hence, the PCRB for $\theta_{i,t}$ at time slot $t$ is given by:
\begin{equation}\label{eq66_1}
\footnotesize
\begin{aligned}
        &PCRB(\theta_{i,t}) = \left(\mathbf{J}^{p}_{i,t} \right)^{-1}_{[1,1]}\\
        &= \left({J}_{\theta_{i,t} \theta_{i,t}} + {\mathbf{M}_{i, t | t-1}^{-1}}_{[1,1]} - \mathbf{J}_{\theta_{i,t} \beta_{i,t}} \mathbf{J}^{-1}_{\beta_{i,t} \beta_{i,t}} \mathbf{J}_{\theta_{i,t} \beta_{i,t}}^T\right)^{-1},
\end{aligned}
\end{equation}
where $\mathbf{A}_{[1,1]}$ means the element of matrix $\mathbf{A}$ in row 1 and column 1.

\begin{remark}
Steps (iv)-(vi) of the EKF procedures are not required to compute the PCRB at time slot $t$. However, for the subsequent time slot, when predicting the MSE matrix, we rely on the computed MSE matrix $\mathbf{M}_t$ obtained at time slot $t$.
\end{remark}

\section{Problem Formulation and Algorithm Design}

\subsection{Problem Formulation}
In designing beamformers and determining the semantic extraction ratio for moving vehicles at time $t$, our objective is twofold: first, to maximise the worst-case SSR among all intended vehicles; second, to enhance the accuracy of vehicle location estimations and tracking performances by minimising the sum PCRB of all vehicles. Consequently, the optimisation problem is formulated as
\begin{subequations}\label{eq66}
\begin{align}
    \max_{\mathbf{W}_{k,t} \succeq 0, \mathbf{R}_{i,t} \succeq 0, \rho_{k,t}}&   \kappa_1 \min_{k\in K}(SSR_{k,t}) - \kappa_2 \sum_{i \in L \cup K} PCRB(\theta_{i,t})\label{eq66a}\\
    \text{s.t.} \quad & \rho_{k,t}^{\text{LB}} \leq \rho_{k,t} \leq 1, \forall k,\label{eq66b}\\
    &  P^{\text{comp}}_t + P^{\text{c\&s}}_t \leq P_t,\label{eq66d}\\
    & \text{rank}\left(\mathbf{W}_{k,t}\right) = 1,  \forall k,\label{eqGOPg}
\end{align}
\end{subequations}
where $\rho_{k,t}^{\text{LB}}$ is given in \eqref{eq14}, $\kappa_1$ and $\kappa_2$ are the trade-off coefficients, and $P_t$ is the total transmit power budget. The notation $\succeq$ stands for positive semi-definite.

The rank-one constraint simplifies beamforming design by ensuring single-stream transmission per user, reducing the need for complex signal-processing techniques. Without this constraint, advanced transceiver designs, such as water-filling power allocation and interference cancellation, or sophisticated signal processing techniques like singular value decomposition (SVD), would be required, significantly increasing computational complexity \cite{wang2014outage}.

\vspace{-0.3cm}
\subsection{Algorithm Design}

Due to mobility, Gaussian errors may be present in the estimated vehicle angles. As a result, the channel of a vehicle can be represented as $\mathbf{h}_{i,t} = \Bar{\mathbf{h}}_{i,t} + \Delta\mathbf{h}_{i,t}$, where $\Bar{\mathbf{h}}_{i,t} = \hat{\beta}_{i, t|t-1} \mathbf{a}\left(\hat{\theta}_{i, t|t-1}\right)$ is the $i$-th vehicle's channel based on state prediction. In addition, $\Delta\mathbf{h}_{i,t}$ is the channel state information (CSI) error vector with zero mean and variance of $\mathbf{\Omega}_{i,t}$, and therefore $\Delta\mathbf{h}_{i,t}$ can be written as: $\Delta \mathbf{h}_{i,t} = \mathbf{\Omega}_{i,t}^{\frac{1}{2}} \mathbf{e}_{i,t}, \quad \mathbf{e}_{i,t} \sim \mathcal{CN}\left(0, \mathbf{I}\right)$.

As mentioned in \cite{wang2014outage}, the rate outage probability constraint is deemed more appropriate than a rate constraint in the presence of Gaussian channel errors. Furthermore, the non-convex constraint associated with the PCRB can be converted into a convex one by introducing a new variable $U_{i,t} \geq 0, i \in L \cup K $. Consequently, optimisation problem \eqref{eq66} is modified to:
\begin{subequations}\label{eqopt}
\begin{align}
    \max_{\mathbf{\Psi}}\; &  \kappa_1 (\lambda - \varrho) - \kappa_2 \sum_{ i \in L \cup K } U_{i,t}^{-1}\label{eqopta}\\
    \text{s.t.} \quad & \Pr(S_{k,t} \geq \lambda) \geq 1 - \epsilon_1, \forall k, \label{eqoptb} \\
    & \Pr(S_{l | k, t} \leq \varrho) \geq 1 - \epsilon_2, \forall l, \forall k, \label{eqoptb2}\\
    &  \begin{bmatrix}
        {J}_{\theta_{i,t} \theta_{i,t}} + {\mathbf{M}_{i, t | t-1}^{-1}}_{[1,1]} - U_{i,t} & \mathbf{J}_{\theta_{i,t} \beta_{i,t}} \\
        \mathbf{J}^T_{\beta_{i,t} \theta_{i,t}} & \mathbf{J}_{\beta_{i,t} \beta_{i,t}}
    \end{bmatrix} \succeq 0, \forall i,\label{eqoptc}\\
    &\eqref{eq66b}, \eqref{eq66d}, \eqref{eqGOPg}, \label{eqoptf}
\end{align}
\end{subequations}
where $\mathbf{\Psi} = [\mathbf{W}_{k,t} \succeq 0, \mathbf{R}_{i,t} \succeq 0, U_{i,t}, \rho_{k,t}, \lambda, \varrho]$. Additionally, $\epsilon_1$ and $\epsilon_2$ are the pre-defined maximum tolerable outage probabilities. For instance, setting $\epsilon_1 = 0.1$ guarantees that $S_{k,t}$ exceeds $\lambda$ at least 90\% of the time.

To tackle the probability constraints, we consider applying the BTI method proposed in \cite{wang2014outage}, which provides computable convex restrictions of the probabilistic constraints. Therefore, by introducing the slack variables $a_k, b_k, a_{l | k}, b_{l | k}$, constraints \eqref{eqoptb} and \eqref{eqoptb2} can be replaced by:
\begin{equation}\label{eq74}
\footnotesize
    \begin{cases}
        \Tr\left(\mathbf{Q}_{k,t}\right) - \sqrt{2 \ln \left(1/\epsilon_1\right)} a_k + \ln\left(\epsilon_1\right) b_k + s_{k,t} \geq 0, \forall k,\\
        \norm{\begin{bmatrix}
            \vect\left(\mathbf{Q}_{k,t}\right)\\
            \sqrt{2} \mathbf{r}_{k,t}
        \end{bmatrix}} \leq a_k, \forall k,\\
        b_k \mathbf{I} + \mathbf{Q}_{k,t} \succeq 0, \forall k, \\
        \vspace{0.2cm}
        b_k \geq 0, \forall k,\\
        \Tr\left(\mathbf{Q}_{l | k,t}\right) - \sqrt{2 \ln\left(1/\epsilon_2\right)} a_{l | k} + \ln\left(\epsilon_2\right) b_{l | k} + s_{l | k,t} \geq 0, \forall k, \forall l,\\
        \norm{\begin{bmatrix}
            \vect\left(\mathbf{Q}_{l | k,t}\right)\\
            \sqrt{2} \mathbf{r}_{l | k,t}
        \end{bmatrix}} \leq a_{l | k}, \forall k, \forall l,\\
        b_{l | k,t} \mathbf{I} + \mathbf{Q}_{l | k,t} \succeq 0, \forall k, \forall l,\\
        b_{l | k} \geq 0, \forall k, \forall l,
    \end{cases}
\end{equation}
\begin{comment}
\begin{equation}\label{eq75}
\footnotesize
    \begin{cases}
        \Tr(\mathbf{Q}_{l | k,t}) - \sqrt{2 \ln(1/\epsilon_2)} a_{l | k} + \ln(\epsilon_2) b_{l | k} + s_{l | k,t} \geq 0, \forall k, \forall l,\\
        \norm{\begin{bmatrix}
            \vect(\mathbf{Q}_{l | k,t})\\
            \sqrt{2} \mathbf{r}_{l | k,t}
        \end{bmatrix}} \leq a_{l | k}, \forall k, \forall l,\\
        b_{l | k,t} \mathbf{I} + \mathbf{Q}_{l | k,t} \succeq 0, \forall k, \forall l,\\
        b_{l | k} \geq 0, \forall k, \forall l,
    \end{cases}
\end{equation}
\end{comment}
where the symbol $\vect ( \cdot) $ is the vectorization of a matrix, and
\begin{equation}\label{eq70}
\footnotesize
    \begin{cases}
    \mathbf{\chi}_{k,t} = \left( \frac{1}{\hat{\gamma}_{k,t}}\mathbf{W}_{k,t} - \sum_{k'\in K, k'\neq k} \mathbf{W}_{k',t} - \sum_{ i \in L \cup K } \mathbf{R}_{i,t}\right), \\
     \mathbf{Q}_{k,t} = \mathbf{\Omega}_{k,t}^{\frac{1}{2}}\mathbf{\chi}_{k,t} \mathbf{\Omega}_{k,t}^{\frac{1}{2}}, \; \mathbf{r}_{k,t} = \mathbf{\Omega}_{k,t}^{\frac{1}{2}}\mathbf{\chi}_{k,t}\Bar{\mathbf{h}}_{k,t}, \\
     \vspace{0.2cm}
     s_{k,t} = \Bar{\mathbf{h}}_{k,t}^H  \mathbf{\chi}_{k,t}  \Bar{\mathbf{h}}_{k,t} - \sigma^2_c, \; \hat{\gamma}_{k,t} = 2^{\lambda \frac{\rho_{k,t}}{\iota}} - 1,\\
     \mathbf{\chi}_{l|k,t} = \left ( \sum_{ i \in L \cup K } \mathbf{R}_{i,t} - \frac{1}{\hat{\Gamma}_{l | k, t}}\mathbf{W}_{k,t} \right),\\
     \mathbf{Q}_{l | k,t} = \mathbf{\Omega}_{l,t}^{\frac{1}{2}}\mathbf{\chi}_{l|k,t} \mathbf{\Omega}_{l,t}^{\frac{1}{2}}, \; \mathbf{r}_{l | k,t} = \mathbf{\Omega}_{l,t}^{\frac{1}{2}}\mathbf{\chi}_{l|k,t} \Bar{\mathbf{h}}_{l | k,t}, \\
     s_{l | k,t} = \Bar{\mathbf{h}}_{l | k}^H  \mathbf{\chi}_{l|k,t}  \Bar{\mathbf{h}}_{l | k,t} + \sigma^2_{c}, \; \hat{\Gamma}_{l | k,t} = 2^{\varrho\frac{\rho_{k,t}}{\iota}} - 1.
    \end{cases}
\end{equation}

As such, the original optimisation problem \eqref{eq66} is transformed to:
\begin{subequations}\label{eq77}
\begin{align}
    \max_{\mathbf{\Psi}}\; &  \kappa_1 (\lambda - \varrho) - \kappa_2 \sum_{i \in L \cup K} U_{i,t}^{-1}\label{eq77a}\\
    \text{s.t.} \quad & \eqref{eqoptc}, \eqref{eqoptf}, \eqref{eq74},\eqref{eq70}
\end{align}
\end{subequations}
where \scriptsize $ \mathbf{\Psi} = [\mathbf{W}_{k,t} \succeq 0, \mathbf{R}_{i,t} \succeq 0, U_{i,t}, \rho_{k,t}, \lambda, \varrho,a_k, b_k, a_{l | k}, b_{l | k}]$. 

\normalsize
To solve \eqref{eq77}, we drop the rank-one constraint and then apply the alternating optimisation. The complete steps for solving \eqref{eq77} are shown in Algorithm \ref{alg:2}.

\begin{algorithm}
\caption{Alternating Optimisation Algorithm for ISCSC}\label{alg:2}
\begin{algorithmic}[1]
\REPEAT
\STATE Conduct EKF procedures (i)-(iii). 
\REPEAT
    \STATE With fixed $\rho_{k,t}^{\text{it}-1}, \lambda^{\text{it}-1}, \varrho^{\text{it}-1}$, solve \eqref{eq77} to obtain $\mathbf{W}_{k,t}^{\text{it}}, \mathbf{R}_{i,t}^{\text{it}}, U_{i,t}^{\text{it}}$.
    \STATE With fixed $\mathbf{W}_{k,t}^{\text{it}}, \mathbf{R}_{i,t}^{\text{it}}$, find $\rho_{k,t}^{\text{it}}$ by using search methods, e.g., the bisection method.
    \STATE Increase $\lambda$ by $\Delta \lambda$ to obtain $\lambda^{\text{it}}$ and decrease $\varrho$ by $\Delta \varrho$ to obtain $\varrho^{\text{it}}$.
\UNTIL{the optimisation problem \eqref{eq77} converges with $\left|\mathbf{W}_{k,t}^{\text{it}} - \mathbf{W}_{k,t}^{\text{it}-1}\right| \leq \varepsilon$ and $\left|\mathbf{R}_{i,t}^{\text{it}} - \mathbf{R}_{i,t}^{\text{it}-1}\right| \leq \varepsilon$.}
\STATE Apply Gaussian randomisation to find the rank-one solutions of $\mathbf{W}_{k,t}^{\text{it}}$ and $\mathbf{R}_{i,t}^{\text{it}}$.
\STATE Conduct EKF procedures by following the steps (iv)-(vi).
\UNTIL{the vehicle leaves the RSU coverage area.}
\end{algorithmic}
\end{algorithm}

The complexity of Algorithm 1 is $\mathcal{O}\left(I_1 \left(N + \log \left( I_2 \right) \right) \right)$, where $I_1$ is the number of iterations for solving \eqref{eq77}, $N$ is the size of the matrices, and $I_2$ is the number of iterations required for the bisection search to converge.

\vspace{-0.2cm}
\section{Numerical Results}
In this section, we present numerical results to assess the efficacy of the proposed design. Our setup assumes that the RSU employs ULAs with half-wavelength spacing, utilising a total of 20 antennas. The noise power is set to -30 dBm, and a total power budget of 20 dBm is used. The trade-off terms are assigned a value of 0.5. The lower bound for $\rho$ is set to 0.65. The centre frequency is set to 30 GHz, the coverage area of the RSU is 100 m, and each block duration is set to 0.02 \cite{liu2020radar,meng2023vehicular}. The initial velocity, distance, and angle for vehicle 1 (unintended vehicle, or eavesdropper) and vehicle 2 (intended vehicle, or legitimate vehicle) are set at $[5^\circ, 55\text{m}, 20\text{m/s}]$ and $[15^\circ, 8\text{m}, 5\text{m/s}]$, respectively. Following from \cite{liu2020radar,meng2023vehicular}, for vehicle 1, we set $Q_1 = \text{diag}\left(0.02^2, 0.2^2, 0.5^2, 0.1^2\right)$ and $Q_2 = \text{diag}\left(1, 6 \times 10^{-7}, 2 \times 10^4\right)$. For vehicle 2, we set $Q_1 = \text{diag}\left(0.02^2, 0.1^2, 0.1^2, 0.1^2\right)$ and $Q_2$ stays unchanged. We set $\epsilon_1 = \epsilon_2 = 0.01$, and $\Delta \lambda = \Delta \varrho = 0.1$. Finally, the value of \(\epsilon\) is set to \(10^{-3}\), and the maximum number of iterations permitted per time slot is limited to 100 \cite{tervo2015optimal}.

\vspace{-0.5cm}
\subsection{Tracking Performance}
\begin{figure}[!t]
    \centering
    \begin{subfigure}{0.24\textwidth}
        \centering
        \includegraphics[width=\textwidth]{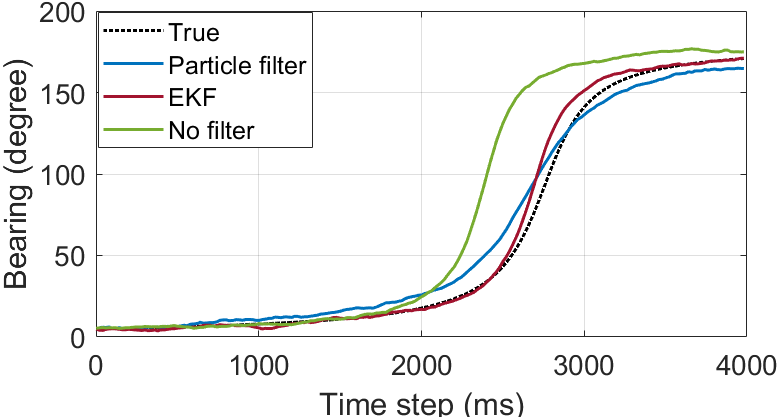}
        \caption{Angle}
        \label{Angle}
    \end{subfigure}
    \begin{subfigure}{0.24\textwidth}
        \centering
        \includegraphics[width=\textwidth]{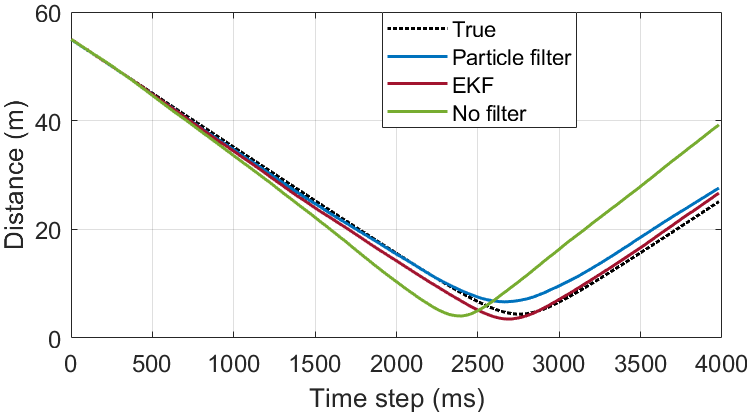}
        \caption{Distance}
        \label{Distance}
    \end{subfigure}
    \caption{Angle and distance tracking performances.}
    \label{tracking performance}
\end{figure}

In this subsection, we only focus on the tracking performance, hence, a simple scenario is considered where a vehicle travels at a constant speed. The tracking results shown in Fig. \ref{tracking performance} demonstrate that using the EKF provides more accurate tracking of both angle and distance compared to scenarios where no EKF is applied or when a particle filter \cite{djuric2003particle, chen2003bayesian} (which is a Monte Carlo-based filter) with 1000 samples is used. In Fig. \ref{tracking performance}(a), around 2500 ms, it is evident that the tracking accuracy of the angle experiences a notable decrease when the vehicle is in front of the RSU without employing the EKF. However, with the implementation of the EKF during this period, the angle prediction closely aligns with the actual values. Fig. \ref{tracking performance}(b) also demonstrates a comparable outcome: the implementation of the EKF results in distance prediction closely matching the actual values. Consequently, the superior tracking performance of the EKF translates into enhanced channel conditions.

\vspace{-0.3cm}

\subsection{Sensing and Communication Performances}
\begin{figure}[!t]
    \centering
    \begin{subfigure}{0.24\textwidth}
        \centering
        \includegraphics[width=\textwidth]{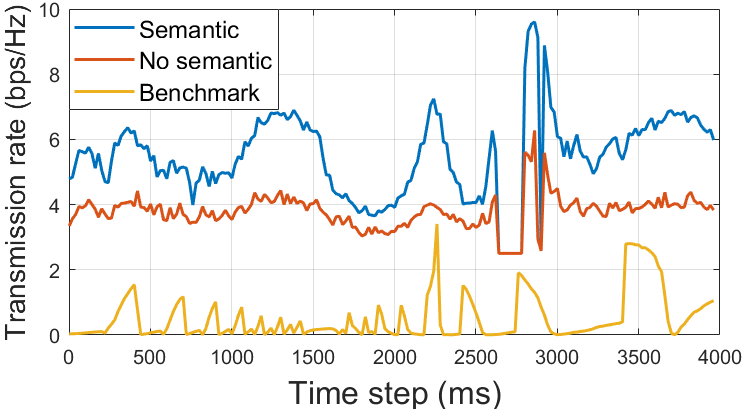}
        \caption{}
    \label{TR_b}
    \end{subfigure}
    \begin{subfigure}{0.24\textwidth}
        \centering
        \includegraphics[width=\textwidth]{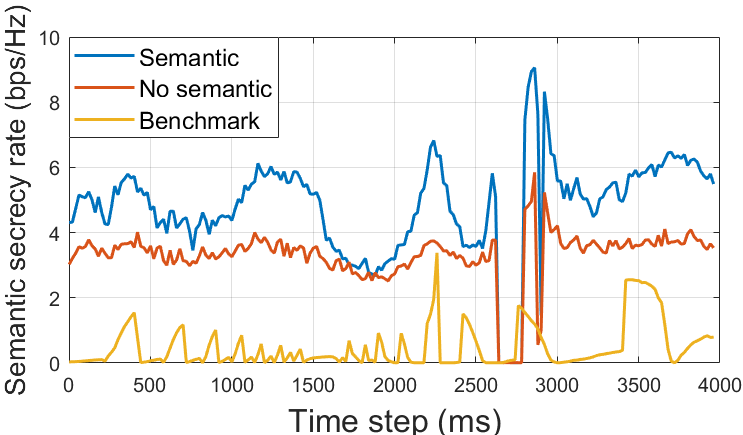}
        \caption{}
    \label{SSR_b}
    \end{subfigure}
    \caption{(a). Transmission rate versus time; (b). Semantic secrecy rate versus time.}
    \label{comm performance}
\end{figure}

As shown in Fig. \ref{comm performance}(a) and Fig. \ref{comm performance}(b), both the transmission rate and the semantic secrecy rate fluctuate due to the mobility of the vehicles. The benchmark for comparison is based on the perfect CSI design, as referenced in \cite{su2022secure}. Our proposed robust design outperforms the benchmark by significantly enhancing the SSR value. On the other hand, when semantic communication is not employed, the average transmission rate is 3.7862 bps/Hz. With the application of semantic communication, this rate increases to 5.4945 bps/Hz, demonstrating notable performance gains. Notably, around 2700 ms, the semantic secrecy rate drops to zero. This can be attributed to the unintended vehicle's proximity to the RSU, allowing it to intercept nearly the same amount of messages as the intended vehicle. Despite the unintended vehicle's presence, the transmission rate to the intended vehicle remains robust throughout the entire journey.

\begin{figure}[!t]
\centering
    \includegraphics[width=0.3\textwidth]{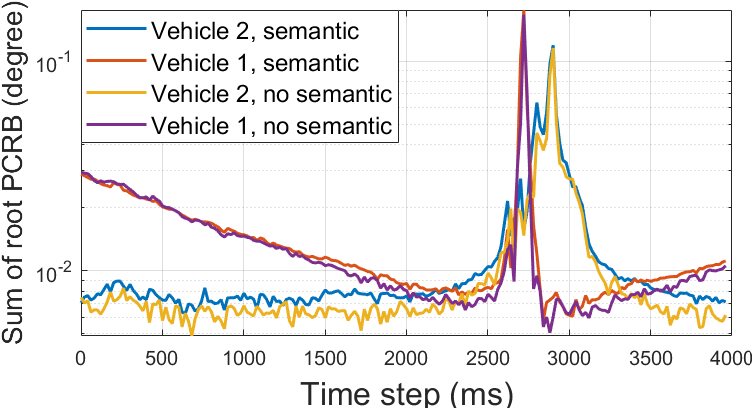}
    \caption{Sensing performance versus time.}
    \label{PCRB_b}
\end{figure}

The sensing performance is illustrated in Fig. \ref{PCRB_b}. Regardless of whether semantic security techniques are employed or not, the sensing performances exhibit minimal differences. However, it is noteworthy that when the vehicles are positioned directly in front of the RSU, the utilisation of semantic techniques leads to a slightly degraded tracking performance compared to the scenario without semantic. Based on the results presented, we can conclude that incorporating semantic communication techniques into the system achieves significantly enhanced security capabilities while maintaining nearly identical sensing performance.

\section{Conclusion}
This paper explores the joint secure design of transmit beamforming vectors and semantic extraction ratios for an ISCSC system involving multiple vehicles. Performance indicators, such as the semantic transmission rate and the semantic computing power, are applied to measure communication and sensing performance. A secrecy outage-constrained algorithm is designed to overcome channel uncertainties. The non-convex optimisation problem is tackled by applying the Bernstein-type inequality and the alternating optimisation methods. Simulation results demonstrate that the proposed framework and algorithm balance the sensing accuracy, communication throughput, and security in vehicular networks.

\bibliographystyle{ieeetr}
\bibliography{bib}

\end{document}